\documentclass[aps, 12pt, final, notitlepage, oneside, onecolumn, nobibnotes, preprintnumbers, nofootinbib,
superscriptaddress,
centertags] {revtex4}
\usepackage{amsmath,amssymb,amsfonts,epsfig,graphicx}
\newcommand{\be}{\begin{equation}}
\newcommand{\ee}{\end{equation}}

\newcommand{\bse}{\begin{subequations}}
\newcommand{\ese}{\end{subequations}}
\newcommand{\bea}{\begin{eqnarray}}
\newcommand{\eea}{\end{eqnarray}}
\newcommand{\ba}{\begin{array}}
\newcommand{\ea}{\end{array}}

\def \th {\theta^{\mu\nu}}

\def\P{Poincar\'e }

\def\nc{noncommutative\ }
\def\ncy{noncommutativity\ }

\begin{document}
\preprint{IPM/P-2008/062 \cr HIP-2008-37/TH\cr
           \eprint{arXiv:0811.3670 [hep-ph]} }
\vspace*{3mm}\title{{ An Arena for Model Building\\ in the   Cohen-Glashow Very Special Relativity}}%
\author{\firstname{M.M}.~ \surname{Sheikh-Jabbari}}
\affiliation{School of Physics, Institute for Research in
Fundamental Sciences (IPM), P.O.Box 19395-5531, Tehran, Iran}
\email{jabbari@theory.ipm.ac.ir} %
\author{\firstname{A}.~ \surname{Tureanu}}
\affiliation {Department of Physics, University of Helsinki and
Helsinki Institute of Physics, P.O.Box 64, FIN-00014 Helsinki,
Finland} \email{anca.tureanu@helsinki.fi}

\begin{abstract}
The Cohen-Glashow Very Special Relativity (VSR) algebra \cite{VSR}
is defined as the part of the Lorentz algebra which upon addition of
$CP$ or $T$ invariance enhances to the full Lorentz group, plus the
space-time translations. We show that noncommutative space-time, in
particular noncommutative Moyal plane, with \emph{light-like
noncommutativity} provides a robust mathematical setting for quantum
field theories which are VSR invariant and hence set the stage for
building VSR invariant particle physics models. In our setting the
VSR invariant theories are specified with a single deformation
parameter, the noncommutativity scale $\Lambda_{NC}$. Preliminary
analysis with the available data leads to $\Lambda_{NC}\gtrsim 1-10\
{\rm TeV}$. This note  is prepared for the Proceedings of the G27
Mathematical Physics Conference, Yerevan 2008, and is based on
\cite{NCVSR}.

\end{abstract}

\maketitle
\section{Introduction and Motivation}

According to Special Relativity (SR)  physical theories and
observables are invariant under the \emph{Poincar\'e group}, that is
the set of Lorentz transformations plus space-time translations.
Mathematically, the \P algebra is the isometry of the $3+1$
dimensional Minkowski space. One may then consider possible
extensions or restrictions of the \P group and study the theories
which are invariant under specific extensions or restrictions. The
maximal extension of the \P algebra is the conformal group
{$so(4,2)$} which  cannot be a symmetry of the particle physics
models even at classical tree level due to the presence of massive
particles. Besides the conformal group, one can extend \P group (or
algebra) by the addition of the discrete symmetries of space and
time inversion {\emph{$P,\ T$}}.

The discrete symmetries, $P$, $T$ and charge conjugation $C$, at low
energy (where QED+QCD is at work) are individually good symmetries
of nature. However, at higher energies, as is built in the particle
physics standard model (SM), the weak interactions do not respect
the parity invariance. Moreover, experiments and observations
confirm that the charge conjugation times parity $CP$, and hence
$T$, are also violated in the strange and b-meson systems. The $CP$
violating parameters in the SM are encoded in the generation mixing
matrices. Theoretically, there is a generally held belief that the
observed $CP$ violation can be traced back to physics beyond the
electroweak scale (beyond SM). On the other hand, although in the
experiments and observations so far we do not have a decisive signal
of Lorentz symmetry violation, it is conceivable that Lorentz
symmetry is not an exact  symmetry at energies above  the
electroweak scale. Cohen and Glashow in their idea of very special
relativity (VSR) \cite{VSR} were in fact seeking a connection
between the two usually thought to be unrelated phenomena, the $CP$
and the possible Lorentz symmetry violation.

The Cohen-Glashow Very Special Relativity (VSR) \cite{VSR} is
defined as symmetry under certain subgroups of \P group, containing
space-time translations and a proper subgroup of Lorentz group
$SO(3,1)$ with the property that when supplemented with parity, $CP$
or time-reversal $T$ it enlarges to the full Lorentz group. In other
words, a theory with VSR symmetry is not strictly Lorentz invariant
and also not parity or time-reversal invariant. As it is seen from
the definition, and will be made explicit in our setup for the
realization of VSR invariant theories, the Lorentz violation and
$CP$ violation are linked together.

Although currently we do not have any observation or experiment
signaling departure from Lorentz symmetry, with the advance of
technologies we will be able to trace such deviations with ever
increasing precision. With the prospect of upcoming experiments
various possible deviations from Lorentz invariance at high energies
have been studied,  both theoretically and phenomenologically (for
an incomplete list see e.g. \cite{{L.V1},{L.V2}}). The problem open
in \cite{VSR} is whether Lorentz and nearly $CP$ invariant theories,
like the Standard Model, could emerge as effective theories from a
more fundamental scheme, perhaps operative at the Planck scale.

One of the very crucial implications of the Lorentz symmetry is in
building physical models based on Lorentz invariant quantum field
theories: In field theories the fields and/or the states in their
Fock space are labeled by  representations of the Lorentz algebra
and in particular particle states are specified by their mass and
spin and they obey the spin-statistics relation. In the formulation
of any theory in which the Lorentz invariance is relaxed, like the
VSR invariant theories, one should re-examine whether one can use
the outcomes of the Lorentz symmetry about the representations of
the matter content of the theory. In the particular case of VSR
subgroups of the Lorentz group, as it will become clear momentarily,
they only admit one-dimensional representations. Being proper
subgroups, the representations of the VSR subgroups of Lorentz are
automatically representations of the Lorentz group, but the
reciprocal is not true. As a result, if we construct a VSR invariant
quantum field theory based on the one-dimensional representations of
the VSR subgroups, when requiring also $P$, $T$ or $CP$ invariance,
although the theory becomes invariant under the whole Lorentz group,
the fact about the one-dimensional representations of VSR does not
change and hence the effective theory would be doomed by its very
poor representation content. This is what we call ``representation
problem'' in the VSR invariant theories.

In view of the above argument and recalling the fact that the
observed elementary particles are neatly classified and described by
the representations of Lorentz group, one possibility for resolving
the ``representation problem'' is that the Lorentz violating terms
(the terms which reduce the symmetry of the Lagrangian to VSR) are
added as perturbations to ordinary Lorentz invariant Lagrangians and
hence the theories have the usual matter content allowed by Lorentz
invariance. However, such a realization of VSR may not be thought of
as a ``fundamental'' or ``master'' theory which leads to Lorentz
invariant theories at low energies; this approach does not provide a
firm theoretical setting for building the VSR invariant theories.

Here we present an alternative way for resolving the
``representation problem'', paving the way for formulating VSR
invariant quantum field theories. This could be achieved noting that
one can include in the picture of symmetries not only the
commutation relations defining an algebra, but also the action of
the generators of the symmetry on the tensor product of their
representation spaces (the so-called co-product). In more
mathematical terms, the reasoning in terms of Lie groups/algebras
can be extended to considering (deformed) Hopf algebras. In the
framework of Hopf algebras, there are deformations which leave the
commutation relations and structure constants of the algebra
untouched, but affect other properties of the Hopf algebra, i.e. the
co-algebra structure \cite{monographs}. Since the commutation
relations of generators are not deformed, it follows automatically
that the Casimir operators are the same and the representation
content of the deformed Hopf algebra is identical to the one of the
undeformed algebra. On the other hand, the deformation of the
co-algebra structure reduces the symmetry of the scheme. Such
deformations are the twists introduced in Ref. \cite{Drinfeld83},
and are hence called ``Drinfeld twist''. Since the twisting reduces
the symmetry, one may try to use the same concept to reduce the
Lorentz symmetry to its VSR subgroups.  This is indeed the idea we
are putting forward here to give a robust mathematical framework for
constructing the VSR invariant theories.

The paper is organized as follows. In section II, we review the
Cohen-Glashow VSR subgroups of Lorentz. In section III, we recall
some aspects about noncommutative space-times and in particular the
Moyal space. In section IV, we show that there are specific
noncommutative space-times, with light-like noncommutativity, that
are invariant under the $T(2)$, $E(2)$ or $SIM(2)$ subgroups of the
Lorentz group. In section V, we present a general framework for
writing $T(2)$ VSR invariant quantum field theory.

\section{The Cohen-Glashow VSR: A brief review}

Energy-momentum conservation, and hence invariance under rigid
space-time translations, should be preserved in VSR invariant
theories. The minimal version of the VSR algebra contains, besides
the generators of translations $P_\mu$, the subgroup $T(2)$ of the
Lorentz group, which is generated by
\be\label{t1t2}%
T_1=K_x+J_y\ \ \ \mbox{and}\ \ \  T_2=K_y-J_x,%
\ee%
where $J_i$ and $K_i$ $i=x,y,z$ are respectively  generators of
rotations and boosts. It is then immediate to check that $[T_1,
T_2]=0$ and hence $T(2)$ is an Abelian subalgebra  of Lorentz
algebra
$so(1,3)$. Moreover, upon action of parity ${P}$,%
\be\label{T2-parity}%
{T_1} \longrightarrow {T_1^P=-K_x+ J_y}\ ,\qquad {T_2}
\longrightarrow {T_2^P=-K_y- J_x}\ ,%
\ee%
and similarly under $T$. It is straightforward to see that the
algebra obtained from $T_1,\ T_2, T_1^P$ and $T_2^P$ closes on the
whole Lorentz group and therefore, $T_1,T_2, P_\mu$ form (the
smallest possible) VSR algebra.

The group $T(2)$ can be identified with the translation group on a
two dimensional plane. The other larger versions of VSR are obtained
by adding one or two Lorentz generators to $T(2)$, which have
geometric realizations on the two dimensional
plane:%
\begin{itemize}%

\item $E(2)$, the 3-parametric group of two dimensional
Euclidean motion, generated by $T_1,\ T_2$ and $J_z$, with the
structure:
\be [T_1,T_2]=0,\ \ [J_z,T_1]=-iT_2,\ \ [J_z,T_2]=iT_1; \ee
\item $HOM(2)$, the group of orientation-preserving similarity
transformations, or homotheties, generated by $T_1,\ T_2$ and $K_z$,
with the structure
\be [T_1,T_2]=0,\ \ [T_1,K_z]=iT_1,\ \ [T_2,K_z]=iT_2; \ee
\item $SIM(2)$, the group isomorphic to the four-parametric
similitude group, generated by $T_1,\ T_2,\ J_z$ and $K_z$.
\end{itemize}
Some comments about the above VSR subgroups are in order:%
\begin{itemize}%
\item It is obvious that once we add parity or time-reversal
conjugates of the generators to either of the above three VSR
cases, similarly to the $T(2)$ case, they enhance to the full
Lorentz algebra.
\item The {$T(2)$} VSR has an {invariant vector}
${n_\mu=(1,0,0,1)}$ as well as {an invariant two form} \cite{VSR},
to which we  shall return  later.
\item ${n_\mu=(1,0,0,1)}$ is also an {invariant vector} of the {$E(2)$} VSR \cite{VSR}.
${E(2)}$ does not have any invariant two form.
\item The ${HOM(2)}$ and {$SIM(2)$}  do not admit any
invariant vector or tensors.
\item The VSR subgroups only admit one dimensional representations.
While all the representations of VSR are also representations of the
Lorentz group, the converse is not true.

\end{itemize}%

One way to realize {$T(2)$} or {$E(2)$} VSR is to recall that they
admit invariant vector or tensor and use the idea of \emph{inverse}
spontaneous Lorentz symmetry breaking and  give VEVs to a vector or
a tensor in such a way that in low energies the VEV goes away, or
become negligible and we recover the full Lorentz symmetry. This was
indeed the idea put forward by {Cohen and Glashow} \cite{VSR, L.V1}
 and some other authors.
However, this may spoil the nice features of Lorentz invariant
theories and in principle is a {phenomenological} approach which
introduces many parameters in the theory.

 ${HOM(2)}$ and {$SIM(2)$} do not admit invariant tensors and their formulation
should be done in some other ways. The {$SIM(2)$} case as the
largest VSR has been studied more (see  e.g. \cite{SIM2-HOM2}).

\section{A brief Review on Noncommutative Spaces and the twisted \P algebras}

As mentioned above, the \P algebra is the isometry of the Minkowski
space. The idea we will follow here is whether there are $3+1$
dimensional space-times whose ``isometry'' group is either of the
VSR subgroups.  Noncommutative spaces which are defined through the
commutation relations
\be\label{cr}%
[x^\mu,x^\nu]=i\th\  %
\ee%
among their coordinates, where $\th$ is in general a function of
coordinates (of course, with the condition that it satisfies the
Jacobi identity), provide a setup to address this issue.

The commutation relations \eqref{cr} usually spoil the Lorentz
invariance (and  the translational invariance if $\theta$ has
space-time dependence). Nonetheless, depending on $\theta$, specific
subgroups of the \P group under which the commutation relation
\eqref{cr} is preserved  still provide a symmetry (or ``isometry'')
of the noncommutative space-time.

The essential element for our discussion is that for specific
choices  of $\theta$  the commutation relations can be obtained from
the associative star-products coming from  introducing twisted
co-product for the \P algebra \cite{CKNT,CPrT} (see also \cite{LW,
{Tureanu:2007zza}}). The advantage of using the twisted \P language
for constructing physical theories is that, in spite of the lack of
full Lorentz symmetry, the fields carry representations of the full
Lorentz group
 \cite{CKTZZ,CNST} and the spin-statistics theorem is still valid; the deformation then
appears in the product of the fields (interaction terms) and
therefore, we have a way of overcoming the ``representation
problem''.

Depending on the structure of the r.h.s. of \eqref{cr}, there exist
three types of noncommutative deformations of the space-time which
can be realized through twists of the \P algebra \cite{CKNT,CPrT,LW}:%
\begin{itemize}
\item
\textit{Constant $\th$}: the Heisenberg-type commutation relations,
defining the \textit{Moyal space}:
\be\label{heisenberg} [x^\mu,x^\nu]=i\th \,, %
\ee
where $\th$ is a constant antisymmetric matrix. This is the most
studied case and various aspects of QFTs on the Moyal space have
been analyzed. Here we briefly review them. Since $\th$ is an
anti-symmetric two tensor, \nc spaces can be classified based on the
values of the two Lorentz
invariants%
\be%
\Lambda^4\equiv \theta_{\mu\nu}\theta^{\mu\nu},\ \ \
   L^4\equiv
   \epsilon^{\alpha\beta\mu\nu}\theta_{\mu\nu}\theta_{\alpha\beta}.
\ee%
$\Lambda^4$ is related to the \ncy scale, the scale where
noncommutativity effects will become important, while $L^4$ is
related to the smallest (space-time) volume that one can measure in
a \nc theory.

Depending on whether $L^4$ and $\Lambda^4$ are positive, zero or
negative one can recognize nine cases. The $L^4\neq 0$ cases cannot
be obtained as a decoupling (low energy) limit of open string
theory. (However, the $\Lambda ^4=0$, $L^4\neq 0$ case is the famous
Doplicher-Fredenhagen-Roberts \cite{DFR} noncommutative space.)

 For $L^4=0$,
depending on the value of $\Lambda^4$, there are three types of
noncommutative spaces:\\ {\it i)} $\Lambda^4>0$ -- space-like
(space-space) noncommutativity;\\ {\it ii)} $\Lambda^4<0$ --
time-like (time-space) noncommutativity;\\ {\it iii)} $\Lambda^4=0$
-- light-like
noncommutativity.\\
When $\Lambda$ is  constant, for the case {\it ii)}, it has been
shown that there is no well-defined decoupled field theory limit for
the corresponding open string theory \cite{AGM}. In the field theory
language this shows itself as instability of the vacuum state
\footnote{This instability is similar to the instability caused by
background electric fields due to pair creation if the theory has
massless charged particles.} and non-unitarity of the field theory
on time-like \nc space \cite{GM}. For the space-like case {\it i)}
and light-like case {\it iii)}, noncommutative field theory limits
are well-defined and the corresponding field theories are
perturbatively unitary.%

\item \textit{Linear $\th$}, with the Lie-algebra type commutators:%
\be\label{lie}%
 [x^\mu,x^\nu]=iC_\rho^{\mu\nu} x^\rho \,. %
\ee%
This case describes an (associative but) noncommutative space if
$C_\rho^{\mu\nu}$ are structure constants of an associative Lie
algebra. %
\item \textit{Quadratic noncommutativity}, with the quantum group type
of commutation relations:%
\be\label{qg}%
[x^\mu,x^\nu]=\frac{1}{q}R_{\rho\sigma}^{\mu\nu} x^\rho x^\sigma \,. %
\ee
\end{itemize}

All the above-mentioned cases of noncommutative space-time have
originally been studied in \cite{CDP} with respect to the
formulation of NC QFTs on those spaces.

\section{Noncommutative Space-times Invariant under the VSR subgroups of Lorentz}

As already pointed out, the VSR  is defined through a proper
subgroup of the Lorentz group $SO(3,1)$, which could be either of
$T(2)$, $E(2)$, $HOM(2)$ or $SIM(2)$, plus space-time translations
generated by $P_\mu$. Among the three cases of NC space-time
discussed in the previous section only the constant $\th$ case
preserves the space-time translational invariance in all directions.
In the cases of linear and quadratic noncommutativity, translational
invariance along some or all of the space-time directions is lost.
Therefore, the Moyal case is the one relevant to the VSR theory. For
completeness, however, we also discuss the relevance of the linear
and quadratic noncommutative cases to the VSR subgroups of Lorentz.

\subsection{$T(2)$ symmetry implies light-like noncommutativity}

Motivated by the above arguments, we set about finding a
configuration of the antisymmetric matrix $\th$. Since $T(2)$ is the
only VSR which admits an invariant two tensor \cite{VSR}, we focus
on this case. If we denote the elements of the $T(2)$ subgroup by
\be\label{l1l2} \Lambda_1=e^{i\alpha T_1}\ \ \ \mbox{and}\ \ \
\Lambda_2=e^{i\beta T_2}, \ee
the invariance condition for the tensor $\theta^{\mu\nu}$ is written
as:
\be\Lambda^{\phantom{i}\mu
}_{i\phantom{\mu}\alpha}\Lambda^{\phantom{i}\nu
}_{i\phantom{\nu}\beta}\theta^{\alpha\beta}=\theta^{\mu\nu},\ \ \
i=1,2,\ee
and infinitesimally:
\be\label{inv_eq} T^{\phantom{i}\mu
}_{i\phantom{\mu}\alpha}\theta^{\alpha\nu}+T^{\phantom{i}\nu
}_{i\phantom{\nu}\beta}\theta^{\mu\beta}=0,\ \ \ i=1,2.\ee
The matrix realizations of the generators $T_1$ and $T_2$ are (see,
e.g., \cite{CH}):
\begin{eqnarray}\label{t1-matrix}
T_1=\left(
\begin{array}{cccc}
0 &-i & 0  & 0 \\
-i & 0 & 0  & i \\
0 & 0  &0 & 0 \\
0 & -i  & 0 & 0
\end{array}
\right),
\end{eqnarray}
and
\begin{eqnarray}\label{t2-matrix}
T_2=\left(
\begin{array}{cccc}
0 &0 & -i  & 0 \\
0 & 0 & 0  & 0 \\
-i & 0  &0 & i \\
0 & 0  & -i & 0
\end{array}
\right).
\end{eqnarray}

Plugging these values into \eqref{inv_eq}
we find the solution%
\be\label{ll-ncy}%
 \theta^{0i}=-\theta^{3i},\ \ \ i=1,2,%
\ee%
all the other components of the antisymmetric matrix
$\theta^{\mu\nu}$ being zero. Note that to obtain the above result
we did not assume any special form for the $x$-dependence of $\th$
and hence this holds for either of the three constant, linear and
quadratic cases. With the above condition on $\th$, we see that $ \Lambda^4=L^4=0$, that is\\
\centerline{{\it Regardless of its space-time dependence, a
light-like $\theta^{\mu\nu}$ is invariant under $T(2)$}.}

One may use the light-cone frame coordinates%
\be\label{light-cone-coor}%
x^\pm=(t\pm x^3)/2,\qquad x^i\,, \ i=1,2.%
\ee%
In the above coordinate system the only non-zero components of the
light-like \ncy \eqref{ll-ncy} is
$\theta^{-i}=\theta^{0i}=-\theta^{3i}$ (and
$\theta^{+-}=\theta^{+i}=\theta^{ij}=0$). In the light-cone
coordinates (or light-cone gauge) one can take $x^+$ to be the
light-cone time and $x^-$ the light-cone space direction. In this
frame, (light-cone) time commutes with the space coordinates. In the
light-cone $(+,-,1,2)$ basis
\begin{eqnarray}\label{theta}
\theta^{\mu\nu}=\left(
\begin{array}{cccc}
0 & 0 & 0  & 0 \\
0 & 0 & \theta  & \theta' \\
0& -\theta & 0  & 0 \\
0 & -\theta'  & 0 & 0
\end{array}
\right).
\end{eqnarray}

\subsection{$E(2)$ and $SIM(2)$ invariant NC spaces}

A constant $\theta^{-i}$ breaks rotational invariance in the
$(x^1,x^2)$-plane and hence larger VSR subgroups are not possible in
the Moyal NC space case. The $E(2)$ invariant case can be realized
in the linear, Lie-algebra type noncommutative spaces and $SIM(2)$
can be realized by quadratic noncommutativity.
\subsubsection{The $E(2)$ case}
$E(2)$ is made up of {$T_1, T_2, J_z$}. {$x^\pm$} are invariant
under {$J_z$}. {$\delta_{ij}$} and {$\epsilon_{ij}$} are the (only)
two invariant tensors under ${J_z}$ while $x^i$ transform as vector
under $J_z$. Therefore, ${\theta^{-i}=\ell\epsilon_{ij} x^j}$ and
${\theta^{-i}=\ell x^i}$ lead to ${E(2)}$ invariant spaces, namely%
\bse\label{E2-theta}%
\begin{align}%
[x^-, x^i] &=i\ell \epsilon^{\ ij} x^j, \\ \mbox{or}\ \cr [x^-, x^i]
&=i\ell x^i.
\end{align}
\ese%
With the above choices, it is evident that the translational
symmetry along $x^\pm$ is preserved while along $x^i$ it is lost.

Instead of $x^i$ coordinates we may work with the cylindrical
coordinates on $x^-, x^1,x^2$ space.
If we denote the radial and angular coordinate on the
$(x^1,x^2)$-plane by $\rho$ and $\phi$,%
\be%
\rho e^{\pm i\phi}=x^1\pm ix^2\ ,%
\ee%
the case (\ref{E2-theta}a) is then described by:
\be%
[x^-,\rho]=0,\quad [\rho, e^{\pm i\phi}]=0, \quad [x^-, e^{\pm
i\phi}]=\pm \lambda e^{\pm i\phi}\,,\ \ \lambda=2l.%
\ee%
Since $\rho$ commutes with both $x^-$ and $\phi$  we can treat it as
a number (rather than an operator). The above space is then a
collection of NC cylinders of various radii and the axes of the
cylinders is along $x^-$. Demanding the wave-functions to be single
valued under $\phi\to \phi+2\pi$, leads to the discreteness of the
spectrum of $x^-$ in units of $\lambda$.

The (\ref{E2-theta}b) case in the cylindrical coordinates takes the
form%
\be%
[x^-, e^{\pm i\phi}]=0,\quad [\rho, e^{\pm i\phi}]=0,\quad [x^-,
\rho]=i\ell\rho\ .%
\ee%
Here we may treat $\phi$  just as a number and work in a basis
where $\rho$ is diagonal. In this basis
$x^-=i\ell\rho\frac{\partial}{\partial\rho}$.

 There is a twisted \P algebra which provides the symmetry for the
case of (\ref{E2-theta}a) while the other case cannot be generated
by a twist \cite{progress}. In the above, $\ell$ and $\lambda$ are
deformation parameters of dimension length.

\subsubsection{The $SIM(2)$ case}

Since $K_z$ acts on $x^\pm$ as scaling (scaling $x^-$ by, say,
$\kappa$ and $x^+$ by $\kappa^{-1}$) while keeping $x^i$ intact, it
is readily seen that it is impossible to find  $\theta^{-i}$ linear
in the coordinates which is invariant under $HOM(2)$. It is,
however, possible to realize $SIM(2)$ (and hence $HOM(2)$, too) with
quadratic $\theta^{-i}$. To have both the $K_z$ and $J_z$ invariant
noncommutative structures, from the above discussions we deduce that
we should take $\theta^{-i}$ which is linear in both $x^-$ and
$x^i$, therefore the two possibilities are
\bse\label{SIM2-theta}%
\begin{align}%
[x^-, x^i] &=i\frac{\xi-1}{\xi+1}\ \epsilon^{\ ij} \{x^-, x^j\}, \quad \xi \in \mathbb{R}\\
\mbox{or}\ \cr [x^-, x^i] &=i\tan\chi\ \{x^-, x^i\}\,,
\end{align}
\ese%
preserving translational symmetry only along $x^+$ (where $\xi$ and
$\chi$ are  dimensionless deformation parameters). For neither of
the above cases there is any twisted \P of the form discussed in
\cite{LW} to provide these commutators \cite{progress}. The case
(\ref{SIM2-theta}b) in the above mentioned cylindrical coordinates
$x^-, \rho, \phi$ takes the familiar form of a quantum (Manin) plane
\cite{Manin} with $x^-$ and $\rho$ being the coordinates on the
Manin plane.

As mentioned above, the Cohen-Glashow VSR requires translational
invariance, which is only realized in the constant $\th$ case,
therefore we continue with the discussion of QFTs on the light-like
Moyal plane, as the VSR-invariant theories. Further analysis of the
linear and quadratic noncommutativity cases will be discussed in a
future work \cite{progress}.

\section{NC QFT on Light-like Moyal plane as VSR invariant
theory}

So far we have shown that a Moyal plane with light-like
noncommutativity is invariant under the $T(2)$ VSR. We  now provide
a prescription for writing VSR invariant QFT for any given
ordinary Lorentz invariant QFT. Our prescription is:\\
For any given QFT on commutative Minkowski space its VSR invariant
counterpart is a noncommutative QFT, NCQFT, which is  obtained by
replacing the usual product of functions (fields) with the nonlocal
Moyal $*$-product (for a review on NC QFTs see \cite{NCQFT-review})
\be\label{Moyal-product-def}%
(\phi*\psi)(x)=\phi(x)\
e^{\frac{i}{2}\th\overleftarrow{\partial_\mu}\overrightarrow{\partial_\nu}}\
\psi(x)\ , %
\ee%
where $\th$ is the constant light-like \ncy matrix given in
\eqref{ll-ncy} or \eqref{theta}. Without loss of generality one may
use the freedom in choosing the direction of the axes in the
$(x^1,x^2)$-plane such that $\theta'=0$ and our VSR invariant theory
is specified with a single deformation parameter $\theta$.

 Due to twisted \P symmetry, the fields carry representations
of the full Lorentz group, but the theory is only invariant under
transformations in the stability group of $\th$, $T(2)$
\cite{CKTZZ,CNST}. Consequently, the NC QFT constructed on this
space possess also the same symmetry \cite{LAG}, as well as twisted
\P symmetry \cite{CKNT,CPrT}.

\section{Discussion and Outlook}

We have given a framework for constructing VSR invariant quantum
field theories. In analogy with the \P  algebra which has the
geometric interpretation of the isometry group of the Minkowski
space, our realization of the VSR subgroups, among other things,
provides a geometric interpretation for these groups, as the
isometry groups of specific ``noncommutative'' space-times, with
\emph{light-like noncommutativity}. In particular, demanding
invariance under space-time translations restricts us to light-like
\nc Moyal plane which is specified by a single deformation
(noncommutativity) parameter. This case realizes the $T(2)$
invariant Cohen-Glashow VSR. Our realization of VSR theory naturally
resolves the ``representation problem'' that, in spite of the lack
of full Lorentz symmetry, one can still label fields by the Lorentz
representations in a consistent manner. For the NC QFTs we can rely
on the basic notions of fermions and bosons, spin-statistics
relation and CPT theorem \cite{CPT,CNT,spin-stat}. However, as shown
in \cite{CPT} for NC QED, $C$, $P$ and $T$ symmetries are not
individually preserved and these symmetries, along with the  full \P
symmetry may be recovered only in the $\th\to 0$ limit (or at
energies much below the \ncy scale).

Through the parameter $\theta$ of the NC QFT realization of $T(2)$
VSR which has dimension length-square we define the noncommutativity
scale $\Lambda_{NC}=1/\sqrt \theta$. To find bounds on
$\Lambda_{NC}$ we need to compare results based on the NC models to
the existing observations and data. These data can range from atomic
spectroscopy and Lamb-shift (see, e.g., \cite{Lamb-shift}) to
particle physics bounds on the electric-dipole moments of elementary
particles. The preliminary analysis leads to $\Lambda\gtrsim 1-10\
{\rm TeV}$. A thorough analysis of obtaining bounds on
$\Lambda_{NC}$ is postponed to a future work \cite{progress}. First
steps in this direction should involve constructing VSR invariant
Standard Model, which could be done along the lines of
\cite{{NCSM},{NCSM-Wess}}.

M.M.Sh-J. would like to thank organizers of the ``XXVII
International Colloquium on Group Theoretical Methods in Physics'',
Yerevan, Armenia, August 2008, where this work was presented as a
plenary talk by the recipient of the 4th Hermann Weyl prize.



\end{document}